# Band engineering aided by topological edge state proximity effects: Inducing anti-chirality in graphene

Ricardo Y. Díaz-Bonifaz and Carlos Ramírez

Departamento de Física, Facultad de Ciencias, Universidad Nacional Autónoma de México, Apartado Postal 70542, Ciudad de México 04510, México

*Corresponding author e-mail address: carlos@ciencias.unam.mx

**Abstract**

In this work we analyze infinite graphene nanoribbons subjected to non-uniform magnetic fields that produce topological domain walls in the quantum Hall regime. We show how the proximity between edge states from neighboring domains modifies the band structure due to the state coupling near the domain walls. The proximity-induced band deformations produce phenomena such as bulk-like dispersion that coexist with Landau levels and valley-polarized current paths. It is shown that edge state coupling can be enhanced by continuously varying the magnetic field between two non-trivial topological phases. The mechanism by which neighboring edge states modify the band structure is addressed by tracking their wave-functions over isolated bands and by analyzing the magnetic confinement potential near the domain wall. By calculating the local current density, we show that the coexistence of topological edge states with bulk-like dispersion can lead to the appearance of anti-chirality, in which co-propagating currents appear in the edges while the rest of the nanoribbon is occupied with bulk states. The appearance of anti-chirality is justified by comparing the proposed non-uniform magnetic field profile with an anti-chiral modified Haldane model.

1. Introduction

Topological edge states (TES) in low dimensional condensed matter systems have been extensively studied in recent years due to their robustness against disorder [1,2]. TES have wave-functions that decay exponentially into the bulk and are often chiral, which forbids backscattering, making them strong candidates to achieve dissipation-less electron transport in low dimensional circuits [3–5]. Topological localization can occur far from the physical borders of a system in multi-domain topological materials, in which the value of a topological invariant changes between different parts of the system. The interphase between two non-equivalent topological phases is called a domain wall (DW). The TES localized near custom-made DWs can be used to produce narrow current pathways with arbitrary shapes [3,5–7], which might lead to the production of circuits in 2D materials.

A simple way to introduce non-trivial topology in a material is by means of an external magnetic field. For instance, in the quantum Hall effect (QHE) a 2D conducting material is subjected to a strong magnetic field and exhibits a quantized resistance when measured transverse to the current flow direction [8]. This phenomenon is closely related to topology as the Chern number, which is an integer topological invariant, indicates the number and orientation of chiral edge states responsible for the quantization [2,9,10]. A topological DW in the QHE regime can be built by means of external electric fields. An example of this occurs in graphene p-n junctions, in which an electric field induces a conductivity that is electron-mediated in one region and hole-mediated in another [11–13]. By combining strong magnetic fields with piecewise-constant electric fields the Fermi energy can be locally modified so that the local Chern number varies within different parts of the 2D system, thus forming DWs [12,14]. Topological DWs formed by Chern number variations can also appear in the context of the quantum anomalous Hall effect, which occurs in intrinsically magnetic materials [15]. These materials can be used to form tailor-made DWs in multilayer heterostructures [16–18] or by spin alignment using magnetic force microscopy [19].

An important aspect of DW localization is that edge states from neighboring domains are close enough to couple and modify their energy spectrum. For instance, Han et. al. proposed an effective interaction model for counter-propagating edge states of neighboring domains that treats them as single particle states coupled by an interaction term that decays exponentially with the distance [20]. This simple model allows to recover the fact that the number of effective channels in a DWs is given as $|C_2 - C_1|$, where $C_1$ and $C_2$ are the Chern numbers in adjacent domains.

The fact that proximity effects can modify the energy spectrum of TES suggests that the band structures of periodic structures containing DWs can be tailored by controlling features like the decay rate of TES and their proximity. In graphene, specifically, band engineering has aimed at creating band gaps that enables its use as a semiconductor [21], as well as the creation of topological flat bands in which long range correlations create a fertile environment for superconductivity, as has been observed in twisted multilayer graphene structures [22,23]. Topological flat band production in graphene has also been achieved by means of pseudo-potentials induced by strain [24–26].

An interesting milestone for band engineering in 2D materials is the production of anti-chirality, which was recently introduced to graphene nanoribbons by Colomés and Franz in terms of a modified Haldane model [27]. In the original Haldane model, a graphene-like structure is described by a tight-binding Hamiltonian in which the next-nearest-neighbor hopping parameters have a complex phase. These phases can be interpreted as produced by a non-uniform magnetic field that averages to zero over any unit cell but breaks time reversal symmetry [28]. The anti-chiral modified Haldane model inverts the sign of the phases acquired by one of the sub-lattices, which can be associated with a different non-uniform magnetic field that also averages to zero over any unit cell. Within a low energy continuum Hamiltonian description, the modified Haldane model is such that this non-uniform magnetic field is introduced as a pseudo-scalar potential that offsets the valleys and produces co-propagating edge state coexisting with bulk bands of a zig-zag graphene nanoribbon (ZGNR), thus achieving anti-chirality [27]. Anti-chiral edge states have been experimentally observed in metamaterial realizations [29,30], and it has been proposed that the effect might appear in transition metal dichalcogenides due to spin-orbit coupling in a similar fashion as in quantum spin Hall effect experiments [27,31,32]. However, up to our knowledge the effect has not been directly observed in real low dimensional solids to this date.

In this work we analyze how band structures can be intentionally modified through TES coupling in graphene nanoribbons with infinite DWs that are translationally invariant in the same direction as the ribbon. In section 2 we discuss how DWs can be formed by non-uniform magnetic fields and discuss the main differences between this approach and the DWs formed by combining electric and magnetic fields. In section 3 we show how the band structure of the nanoribbon is altered by the interaction of TES in multi-domain ZGNRs. The relation between the band structure and TES proximity effects is confirmed by tracking the eigenmodes from isolated bands. The mechanism by which TES coupling depends on the wave-vector is addressed in section 4 by analyzing the $k-$dependence of the magnetic confining potential. By performing current density calculations, it is shown in section 5 that the coupled TES in DWs produce currents that counter-propagate with respect to the original TES that form the coupled state. We then show that the non-uniform magnetic field profile can be engineered to match the symmetries of a modified Haldane model to produce anti-chirality, which was confirmed by local current density calculations.

## 2. Domain wall creation with non-uniform magnetic fields

In the following we deal with DWs built from non-uniform magnetic fields in the QHE regime. To justify this approach, which differs from the experimentally convenient p-n junction technique, let us consider an infinite 2D material subjected to a magnetic field that produces a magnetic flux $\Phi$ per unit cell. The energy spectrum of the system as a function of $\Phi$ is outlined by the well-known Hofstadter butterfly pattern [33,34]. Due to the connection between the Hall resistance and the Chern number, the Hofstadter pattern can be used to determine the Chern number for any combination of $\Phi$ and energy $E$, working as a topological phase diagram in the QHE regime [35]. In the p-n junction approach, a material is subjected to a strong magnetic field and a local gate potential $V_G$ is added to offset the energy between the gated and ungated areas. Let us think, for instance, that the local Chern numbers are $C_1 = 1$ and $C_2 = -1$ respectively. If the electric field varies continuously in the interphase between the gated and ungated areas, the energy offset also evolves continuously. As $\Phi$ remains constant throughout the sample, by varying the electric field the local topological phase can be tracked across the DW by following the vertical double arrow shown in Fig. 1, where we can see that if the field varies slow enough a local ordinary insulating phase appears near the DW. As this intermediate local phase is topologically trivial, no additional states appear and the TES of the non-trivial areas are separated from each other, suppressing their interactions.

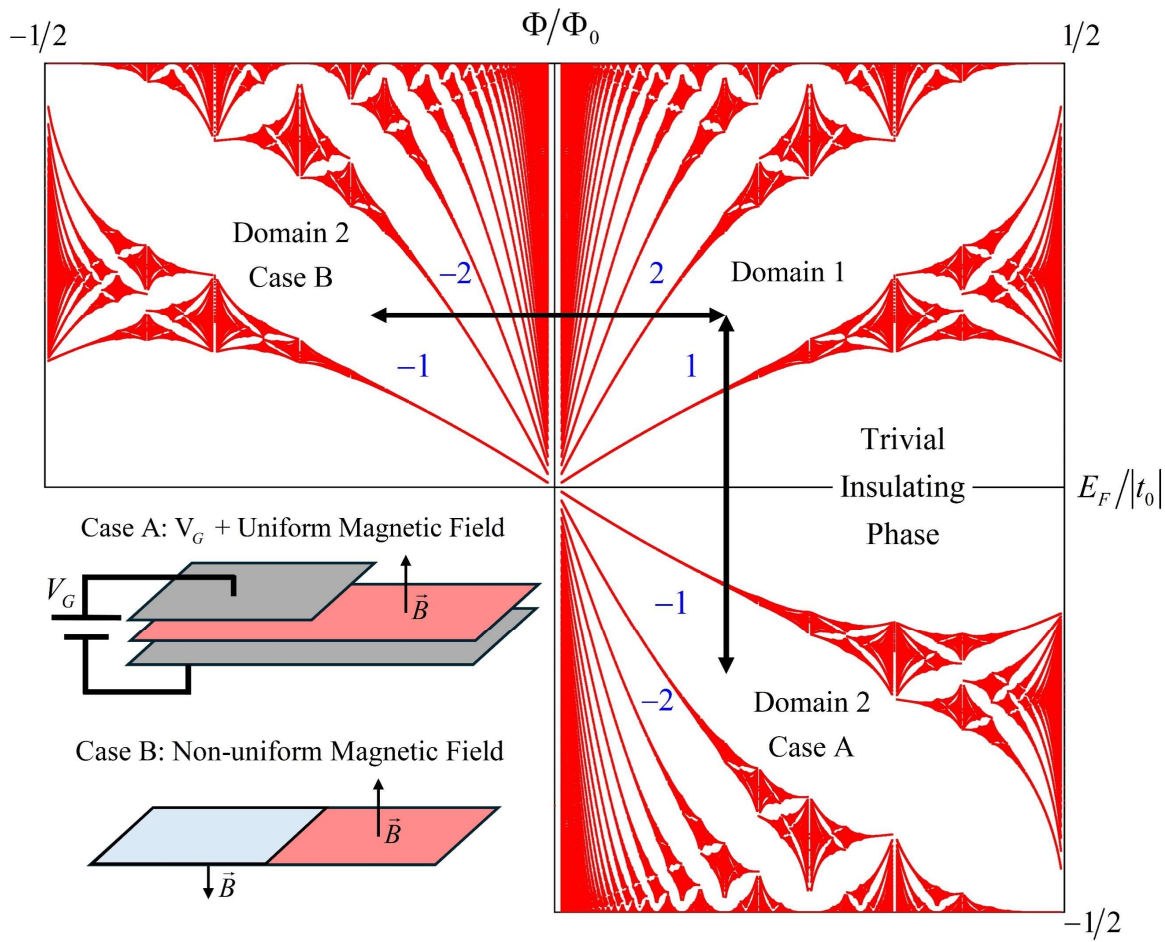

**Figure 1:** Hofstadter pattern (red) that outlines the energy spectrum of a 2D material that has a magnetic flux $\Phi$ per unit cell. The magnetic flux is given in terms of $\Phi_0 = h/e$ while the energies $E$ are given in terms of the hopping parameter amplitude $|t_0|$. The blue numbers indicate Chern numbers for different combinations of $\Phi$ and $E$. Case A is a schematic representation of a Chern DW formed by combining uniform magnetic field with a local gate potential $V_G$. Case B corresponds to a Chern DW formed only by a non-uniform magnetic field. In case A any possible intermediate domain will have a Chern number within the vertical double arrow, while for case B the possible intermediate domains will have Chern numbers are contained in the horizontal double arrow.

Let us now consider a DW formed by a non-uniform magnetic field that produces a flux $\Phi_1$ per unit cell in domain 1 and $\Phi_2$ in domain 2. If the magnetic fields are opposite between domains, we can think of the case where the local Chern numbers are $C_1 = 1$ and $C_2 = -1$, like in the p-n junction example. However, notice that in this case there is no offset in the energy throughout the sample. If the non-uniform magnetic field varies continuously, the local topological phases across the domains follow the horizontal double arrow shown in Fig. 1. Notice that in this case, a smooth evolution of the magnetic field creates new intermediate domains, resulting in additional states whose presence might promote their coupling. As the local Chern number grows when the magnetic field goes to zero for a given energy, we can think of the DW formed by magnetic fields as having a metallic like-behavior which is drastically different from the intermediate insulating phase produced by varying the electric field.

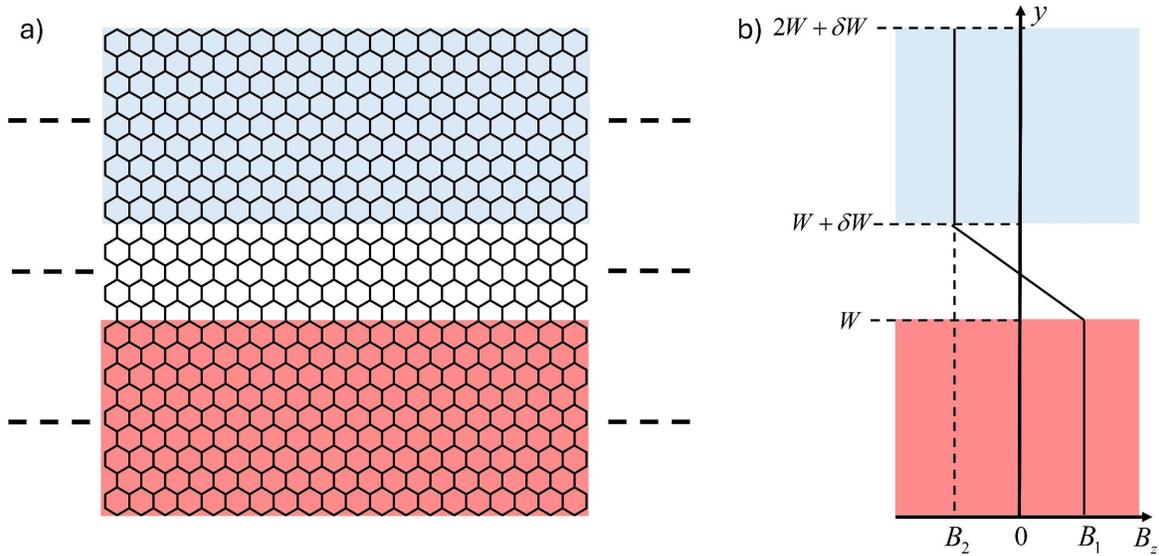

**Figure 2:** a) Schematic representation of a ZGNR infinite and periodic in the $x-$direction where a non-uniform magnetic field produces a constant magnetic flux $\Phi_1 = 3\sqrt{3}a^2 B_1/2$ in the lower domain (red) and $\Phi_2 = 3\sqrt{3}a^2 B_2/2$ in the upper domain (blue), with $a$ being the graphene lattice parameter. $B_1$ and $B_2$ are magnetic field components in the $z-$direction. b) $B_z$ component of the magnetic field as a function of $y$. $B_z$ evolves linearly from $B_1$ to $B_2$ in the portion of the nanoribbon with no colored background.

A periodic 1D structure containing DWs produced by non-uniform magnetic fields can be achieved using infinite nanoribbons. In the following we consider infinite zig-zag graphene nanoribbons (ZGNRs), as their band structures are well documented [36,37], and the band deformations produced by TES proximity effects can be easily identified. The nanoribbons are subjected to a non-uniform magnetic field of the form $\vec{B} = B_z(y)\hat{z}$ that is constant in the $x-$direction, in which the nanoribbon is translationally invariant. In the $y-$direction, the magnetic field profile is such that we can identify a lower domain and an upper domain. In the lower domain, the magnetic field has a constant value of $B_1$ that produces a uniform magnetic flux per unit cell $\Phi_1$, while the upper domain is subjected to an uniform magnetic field $B_2$ that produces a magnetic flux $\Phi_2$ per unit cell. Both domains have the same width $W$. The domains are separated by a $\delta W$ wide section of the nanoribbon where the magnetic field is given as

$$B_z(y) = B_1 + (B_2 - B_1)\frac{y - W}{\delta W}, \tag{1}$$

which results in a linear evolution between the magnetic fields of the lower and upper domains. The nanoribbons are described by a nearest-neighbor tight-binding (TB) Hamiltonian

$$\hat{H} = \sum_{\langle i,j \rangle} t_{i,j} \hat{c}_i^\dagger \hat{c}_j, \tag{2}$$

where $\hat{c}_i^\dagger$ and $\hat{c}_i$ are respectively creation and annihilation operators in the $i-$th lattice site. The hopping parameters are given as $t_{i,j} = t_0 \exp(i\theta_{i,j})$, with $t_0 = -2.7\ eV$ for graphene [36]. Magnetic fields are introduced by means of Peierls phases $\theta_{i,j}$, whose sum over any closed path must be proportional to the net magnetic flux enclosed by such a path [38]. For the proposed non-uniform magnetic field, the Peierls phase assignation is made using the graphic algorithm described in Ref. [39]. The details of the Peierls phase calculations are shown in detail in Appendix A.

### 3. Topological edge state proximity effects

In this section we present the band structures calculated for the ZGNRs defined in the previous sections under different magnetic field profiles. To highlight the bands hosting TES we calculate the inverse participation ratio of every state as

$$IPR_{n,k} = \sum_{i=1}^{N} |\psi_{n,k}(i)|^4, \tag{3}$$

with $N$ being the total number of sites in the ZGNR unit cell and $\psi_{n,k}(i)$ being the normalized wave function coefficient in the $i-$th lattice site for the $n-$th eigenmode of the periodic Hamiltonian $\hat{H}(k)$. In this context, the $IPR_{n,k}$ gives a measure of the localization of any eigenstate in the $y-$direction, in which the ZGNR is finite. A fully delocalized eigenmode is such that $IPR_{n,k} \simeq 1/N$, while a localized eigenmode has $IPR_{n,k} \approx 1$ [40]. For every state with an $IPR_{n,k} \geq 10N^{-1}$, we determine the distance between the localization center and the ZGNR center in the $y-$direction by calculating what we call the *edge confinement* $(EC_{n,k})$, which can be calculated as

$$EC_{n,k} = \sum_{i=1}^{N} |\psi_{n,k}(i)|^2 \frac{|y_i - y_C|}{y_C}, \qquad (4)$$

where $y_i$ and $y_C$ are the $y-$coordinates of the $i-$th lattice site and the nanoribbon center respectively. Within this framework, an $EC_{n,k} \approx 1$ indicates that the state is localized near the (outer) physical borders of the nanoribbon while an $EC_{n,k} \approx 0$ is associated with states localized in the center of the nanoribbon, where the DWs are created.

To identify the proximity effects we calculate the band structure of the ZGNRs when the magnetic fields in the lower and upper domains are respectively $B_1 = 20T$ and $B_2 = -20T$, with both domains having widths of $W = 100$ nm. We consider the cases where the DW is abrupt $(\delta W = 0)$ and where the magnetic field evolves continuously for $\delta W = 30$ nm between domains. These band structures, which can be seen in Fig. 3, are compared with the spectrum of a 100 nm wide nanoribbon subjected to a uniform magnetic field of $B = 20T$.

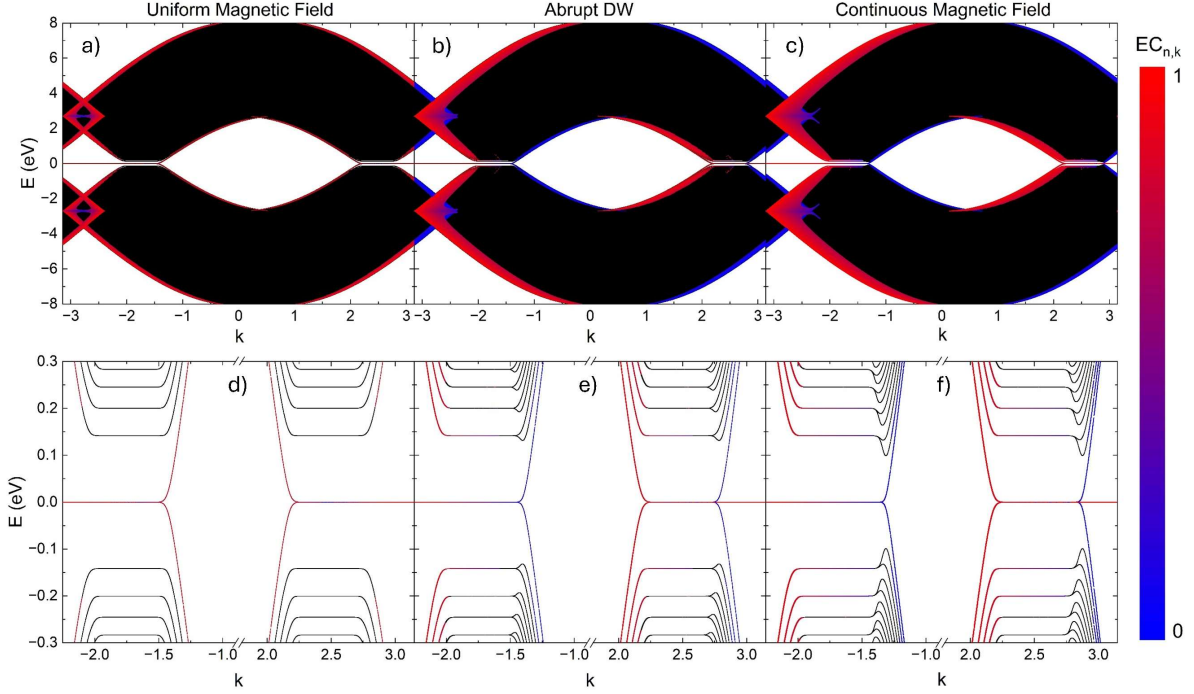

**Figure 3:** Band structure of the ZGNRs shown in Fig. 2. The following parameters are considered. a,d) $B_1 = B_2 = 20T$, $W = 50$ nm and $\delta W = 0$. b,e) $B_1 = -B_2 = 20T$, $W = 100$ nm and $\delta W = 0$. c,f) $B_1 = -B_2 = 20T$, $W = 100$ nm and $\delta W = 30$ nm. d,e,f) are zoomed portions of a,b,c). In all cases the states with $IPR_{n,k} \geq 10 N^{-1}$ are selected for edge confinement $(EC_{n,k})$ calculation, where red means that the state is localized in the physical borders of the ZGNR and blue means that the state is localized at the center.

We start the analysis by pointing out the well-known fact that the uniform magnetic field introduces Landau levels that can be identified as flat bands near the original Dirac points [36], as can be seen in Figs. 3a and 3d. Once the DW is introduced, the TES near the physical borders of the ZGNR (outer TES) propagate in the same direction, while the TES near the DW (inner TES), which originally counter-propagate with respect to the outer TES, are close to each other. The TES located near the center of the ribbon are shown in blue in Figs. 3b, 3c, 3e and 3f. Notice how the left side of the deformed Dirac points in Fig. 3e are built from degenerate bands associated with outer TES that form Landau level flat-bands. On the other hand, on the right side of the deformed Dirac point formed by inner TES this degeneracy is lost and the original Landau levels break into separate bands that form a bulk-like dispersion pattern. The effect is even more evident when the magnetic field evolves continuously for $\delta W = 30$ nm between the domains, as can be seen in Fig. 3f. This coexistence of co-propagating TES with bulk-like dispersive states is further explored in section 5 to produce anti-chirality.

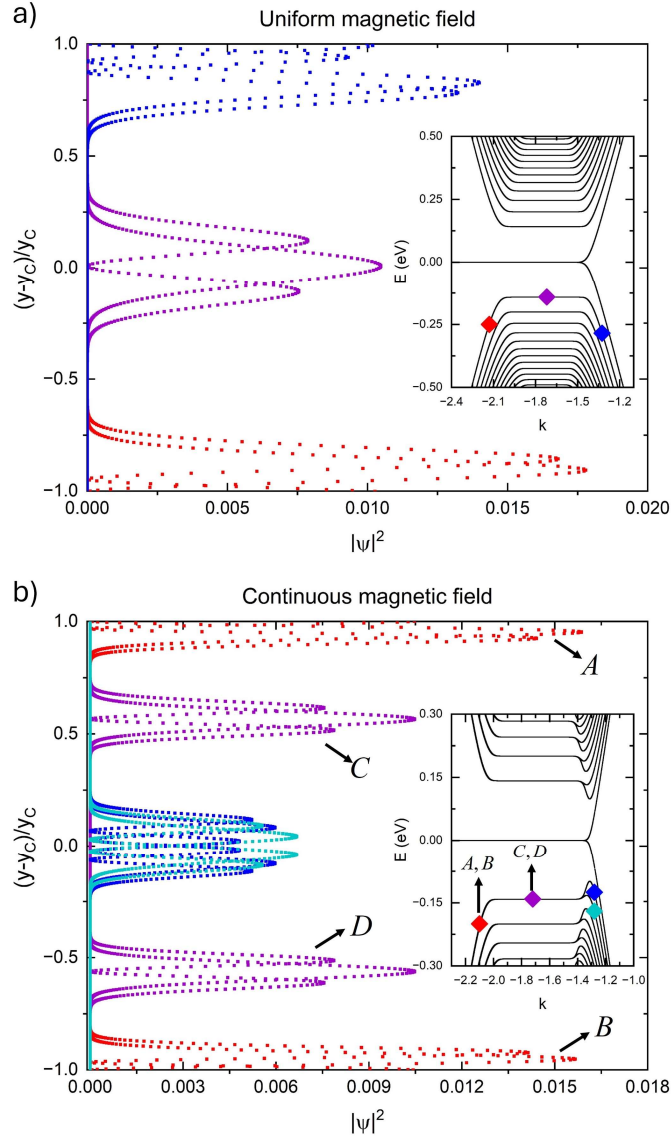

**Figure 4:** a) Squared absolute value of wave-functions in a ZGNR with a constant magnetic field $\left(W = 50 \text{ nm}, B_1 = B_2 = 20T, \delta W = 0\right)$. b) Wave functions in a ZGNR with opposite magnetic fields of a continuous evolution of the magnetic field $\left(W = 100 \text{ nm}, B_1 = -B_2 = 20T, \delta W = 30 \text{ nm}\right)$. In both cases, the color code indicates the states within the band structures shown in the insets. In b) the red and purple diamonds in the inset are two-fold degenerate each, the corresponding wave functions are labeled A,B for the red curves and C,D for the purple curves.

The perturbations in the energy spectra of inner TES can be understood as a proximity effect between those states. To prove this, we calculate the wave-functions of the ZGNR unit cell for different parts of the band structure and show how the deformed bands correspond to coupled TES while the degenerate undeformed bands correspond to isolated TES. Let us start by addressing the ZGNR subjected to a uniform magnetic field. Observe in Fig. 4a that by tracking a TES from the left to the right side of the deformed Dirac point we can see how the wave function is displaced from one edge to the other. For positive group velocities the TES is localized in the lower edge, the flat band section allows the state to be transferred from one edge to the other while keeping the same localization profile, which finally results in a TES that has the opposite group velocity in the opposite border. A similar analysis can be made for the ZGNR with the continuous magnetic field between domains. Notice from Fig. 4b how when the edge states are degenerate, two uncoupled TES can be identified (one in each edge). As we follow these states throughout the flat band we can see how they are transported into the ZGNR center. Observe that as the TES get closer to each other and their wave-functions begin to couple, the band is deformed and the degeneracy is lost, confirming this effect is due to proximity between TES. A continuous visualization of the $k-$parametrized wave-function evolution can be found in the gifs "*Continuous_Field_wave-functions.gif*" and "*Continuous_magnetic_field_wave-functions.gif*" presented as supplemental material. The displacement of edge states as a function of $k$ can be understood by analyzing the confinement potentials induced by non-uniform magnetic fields. This approach is addressed in section 4, showing that these effects are not ubiquitous to ZGNRs.

Another key aspect of the band structures modified by proximity effects occurs in the low energy regime. Notice from Fig. 3d that for a uniform magnetic field, the zero-mode flat band between the Dirac points results in counter-propagating edge states that appear in opposite borders of the ZGNR and have different valley polarization. However, as soon as the DW is introduced (see Figs. 3e and 3f), additional copies of the zero-mode bands appear which now result in co-propagating edge states that have the same valley polarization and their counterparts with opposite group velocities are both localized near the ZGNR center. This means that we now have valley-polarized current pathways as the edges host co-propagating states from the same valley and the bulk is occupied by states from opposite valleys. The phenomenon can be understood in terms of sub-lattice symmetry as in a ZGNR each edge is formed from a different sub-lattice [36], resulting in Landau zero-modes that are supported in different sub-lattices for each edge and possess different valley index. However, the introduction of the DW breaks sub-lattice symmetry which allows both TES to occupy the same valley. Their counterparts, however, are localized near the DW and are coupled, which make them not sub-lattice polarized and thus valley symmetric.

4. **Analysis of magnetic confining potentials**

In this section we show that the relation between the coupled TES and the $k$ vector can be understood in terms of the magnetic confinement potential. Let us consider a two-dimensional electron gas (2DEG) subjected to an non-uniform magnetic field given as $\vec{B}=(0,0,\text{sign}(y)B_0)$, which means that the field for $y>0$ is opposite to the field for $y<0$.

Such magnetic field can be obtained from a vector potential given as $\vec{A} = (-B|y|, 0, 0)$, and can be introduced into the momentum $\vec{p}$ of the free particle Schrödinger equation as $\vec{p} \rightarrow \vec{p} - e\vec{A}$, where $e$ is the electron charge. This results in a Hamiltonian $\hat{H} = \frac{1}{2m}\left((\hat{p}_x - e\vec{A})^2 + \hat{p}_y^2\right)$ that leads to the eigenvalue equation

$$\hat{H}\psi = \frac{1}{2}m\omega_C^2\left(|y| - \frac{\hbar k_x}{m\omega_C}\right)^2 \psi - \frac{\hbar^2}{2m}\frac{\partial^2 \psi}{\partial y^2} = E\psi, \quad (5)$$

where we have introduced the cyclotron frequency $\omega_C = eB/m$ and $\hat{p}_y = -i\hbar\frac{\partial}{\partial y}$ is the real space representation of the $y$–component of the momentum operator. As the Hamiltonian does not depend explicitly on $x$, the $\hat{p}_x$ operator was written in terms of its eigenvalues as $\hat{p}_x \psi = \hbar k_x \psi$. By defininig $y^* = y\sqrt{2m\omega_C/\hbar}$, $k^* = k_x\sqrt{\hbar/2m\omega_C}$ and $E^* = E/\hbar\omega_C$, Eq. (5) can be adimensionalized as

$$-\frac{d^2\psi}{dy^{*2}} + \frac{1}{4}\left(|y^*| - 2k^*\right)^2 \psi = E^*\psi, \quad (6)$$

From Eq. (6) we can identify a $k^*$–dependent magnetic confinement potential defined as $V^*(y) = \frac{1}{4}\left(|y^*| - 2k^*\right)^2$. For positive values of $k^*$, $V^*(y)$ can be interpreted as a pairs of harmonic potentials centered at $\pm 2k^*$. As $k^*$ approaches zero, the harmonic potentials get closer to each other to ultimately collide into a single potential well for $k^* < 0$ where the minimum is $(k^*)^2$. These potentials are shown in the right side of Fig. 5. This analysis explains why by varying $k^*$ the localized states are displaced further or closer to the DW in our previous examples.

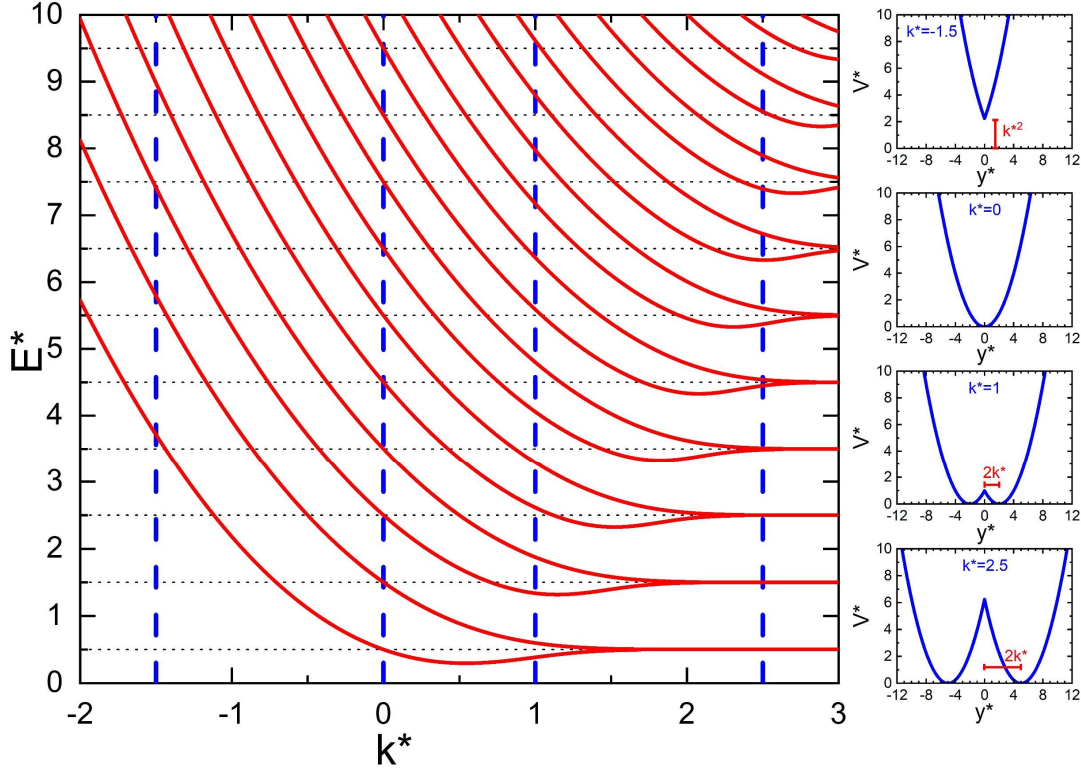

**Figure 5:** Numerical solutions of the adimensional energy $E^*$ from Eq. (6), which explain the deformed bands shown in Fig. 3. In the right side the magnetic confining potential is shown for different values of $k^*$.

The differential equation (6) can be solved for the adimensional energy $E^*$ keeping $k^*$ as a parameter to determine the dispersion relation induced by the non-uniform magnetic field. In Fig. 5 we show the bands obtained when Eq. (6) is numerically solved using the S-matrix based method of Ref. [41]. Observe that when $k^*$ is positive and far from zero, we obtain degenerated energies that form flat bands. This is the limit for which the potential is formed by two independent harmonic oscillators, so the flat band energies correspond to degenerate Landau levels [42]. As $k^*$ approaches zero from the right, the harmonic potentials centers get closer to each other and the eigenfunctions begin to couple which breaks the degeneracy of $E^*$. For $k^* = 0$ we have a single harmonic potential, resulting in an equally spaced energy spectrum. Notice that for $k^* < 0$, where the magnetic potential is formed by a single well the non-degenerate energy spectrum is displaced to higher energies as the well minimum grows as $\left(k^*\right)^2$. This overall behavior is consistent with the band structure obtained for the ZGNR with an abrupt DW (see fig. 3e), which further confirms that the band deformations are produced by TES proximity effects. Moreover, the bulk-like behavior of coupled TES can now be understood as the magnetic confining potential for $k^* < 0$ is that of a single potential well.

## 5. Current density and anti-chirality

As discussed in section 3, the energy spectra of TES are modified due to their proximity, breaking their degeneracy as they form pairs of coupled states that are spectrally isolated from each other. In a periodic setup, like the ZGNRs that we address in this work, the coupled states are dispersive and can propagate with positive and negative group velocities, as opposed to the original TES which were one-directional. To further explore this behavior, we calculate the inter-cell local current density (LCD) $\mathbf{J}_{i,j}^{n,k}$ between every pair of nearest-neighbor sites $i$ and $j$ from adjacent cells of the ZGNR. The LCDs are calculated from the wave-functions as [43]

$$\mathbf{J}_{i,j}^{n,k} = \frac{2}{\hbar}\left(\mathbf{R}_i - \mathbf{R}_j\right) \mathrm{Im}\left(\psi_{n,k}(j)\psi_{n,k}^{*}(i)t_{j,i}\right), \quad (7)$$

where $\mathbf{R}_i$ and $\psi_{n,k}(i)$ are respectively the position vector and wave function amplitude of the $i-$th lattice site for the $n-$th eigenvalue of $\hat{H}(k)$.

We calculate the LCDs for a ZGNR where the magnetic field evolves continuously between domains, which has the band structure shown in Fig. 3f. Observe in Fig. 6c that for an energy of $-0.05\ eV$ there are two degenerate TES localized in the outer borders (red lines) and two eigenstates in the nanoribbon center (blue lines). It can also be seen from the band structure that the inner and outer states have group velocities of different sign. This is consistent with the LCD calculations shown in Fig. 6a, as we can see that the outer border TES produce currents that counter-propagate with the center states. Another interesting feature of these currents is that both outer border TES have energies within the same valley, while the inner states correspond to different valleys despite being equally distributed along the $y-$direction in real space. This behaviour is consistent with the valley-polarized current pathways discussed in Section 3. It can also be seen in Fig. 6b that by choosing an energy of $-0.125\ eV$, which captures the currents produced by coupled edge states in the ZGNR currents, that we now have counter-propagating currents in the ZGNR center, which confirms that the TES proximity effect produces anomalous bidirectional transport from originally co-propagating TES.

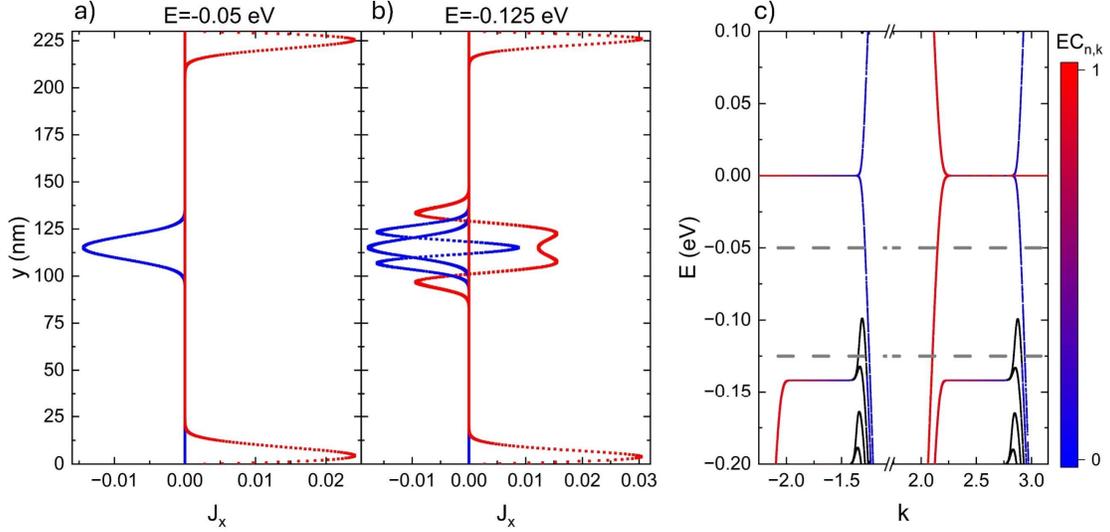

**Figure 6:** Inter-cell local current densities (LCDs) along the transverse section of a ZGNR with 100 nm wide domains with magnetic fields of $B_1 = -B_2 = 20T$ separate by a $\delta W = 30$ nm wide transition region in which the magnetic field evolves linearly between domains. The LCDs are normalized so that each eigen-mode integrates to a unitary value. The LCDs are calculated for all the available eigenmodes with energies of $-0.05\ eV$ for a) and $-0.125\ eV$ for b). The energies for which the LCDs are calculated are shown in c) as dashed lines within the band structure of the ZGNR.

An interesting feature of the LCD profile shown in Fig. 6a is that, as it is formed by two parallel currents in the edges with an opposite current in the center, which is a fingerprint of anti-chirality. For these currents to resemble the anti-chiral behavior reported in previous works [27,44], we must find a magnetic field profile that produces a current that occupies the entire bulk of the ZGNR and counter-propagates with the currents in the outer borders. To achieve this, let us analyze how the anti-chiral Haldane model (AHM) could be reproduced in terms of magnetic fields. This analysis, which is shown in detail in Appendix B, exhibits that the AHM can be produced by subjecting a ZGNR to a periodic magnetic field that averages to zero over any unit cell but is non-uniform within the unit cell and breaks both time-reversal and sub-lattice symmetry. For instance, as it is shown in Appendix B, this non-uniform magnetic field can be defined in a way that the edges of the ZGNR can be associated with opposite local magnetic fields, as occurs in the ZGNRs hereby proposed. It is also worth mentioning that demanding the magnetic field to average to zero over a unit cell allows the appearance of TES without Landau levels [28], which is not within the goals of this work and is not a necessary condition for our model. Thus, to obtain anti-chirality we can think of a ZGNR subjected to a non-uniform magnetic field that evolves continuously throughout the entire nanoribbon width following Eq. (1), which ensures that every unit cell of the ZGNR is subjected to a magnetic field that produces anti-chirality in the same fashion as the AHM.

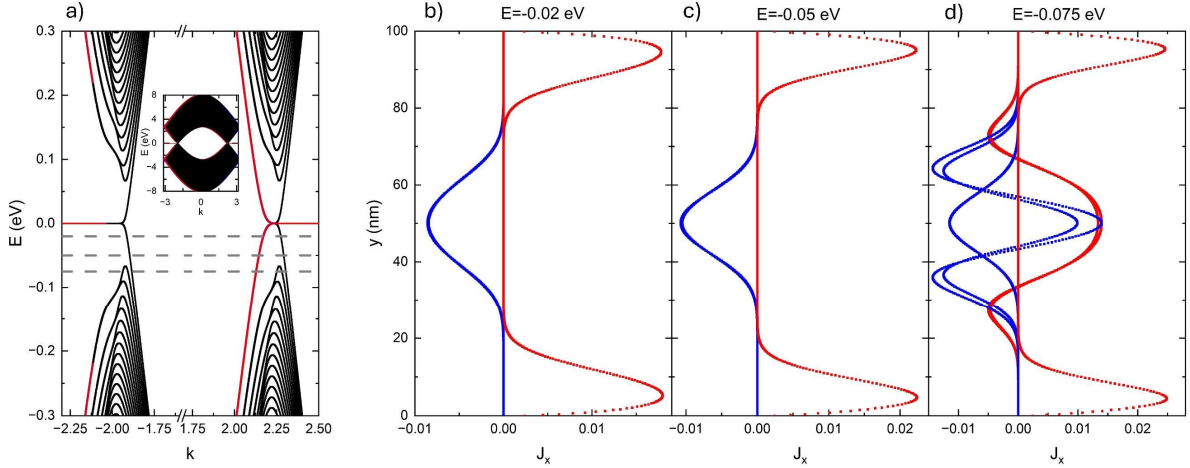

**Figure 7:** a) Band structure of a ZGNR subjected to a magnetic field that evolves linearly for the whole nanoribbon width $(W=0,\ B_1=-B_2=20T,\ \delta W=100\ \text{nm})$. b-d) LCD profiles for different energies, which are marked by dashed gray lines in the band structure. The right-flowing currents are shown in red while left-going currents are shown in blue.

In Fig. 7a we show the band structure of a 100 nm wide ZGNR subjected to a magnetic field that evolves linearly from $B_1=20T$ in the lower edge to $B_2=-20T$ in the upper edge. Notice that in this case the Landau levels that originally formed flat bands are now deformed into dispersive bands that are degenerated in left side of each valley. As in previous cases, this degeneracy is lost as the corresponding states get closer to each other and couple to form a new pair of spectrally isolated states, as can be seen in the wave-function tracking gif "*Anti-chiral_wavefunctions.gif*" included as supplemental material. As a result of these band deformations, we get a bulk-like dispersion that coexists with degenerate bands of co-propagating TES localized in each edge of the nanoribbon, thus achieving anti-chirality. This can be explicitly observed by calculating the LCD, as shown in Figs. 7b-d). Notice that in a low energy limit (see Figs. 7b-c) the resulting LCD profile is formed by co-propagating edge states and counter-propagating states that occupy the rest of the nanoribbon. As we take energies further from zero, the LCD becomes more localized. It is worth mentioning that once again the outer edge states come from the same valley while the counter-propagating bulk states are each from a different valley, which is consistent of them being distributed over both sub-lattices in contrast with the outer edge states which are sub-lattice polarized. Notice as well from Fig. 7d) that if we take an energy such that we capture the coupled states we now get that the bulk has contributions that propagate in both direction. As global Fermi energy can be tuned by a gate potential, we conclude that this mechanism can produce tunable anti-chirality such that the bulk states counter-propagate with the edges and are one-directional or such that they flow in both directions.

## Conclusions

In this work we have analyzed ZGNRs with infinite periodic DWs produced by non-uniform magnetic fields. We have shown that the coupling between TES modify the band structure of the ZGNR. Specifically we found that the coupling of neighboring TES break their degeneracy and the coupled states have a bulk-like dispersion that coexist with Landau levels. The modified band structure of the ZGNR is such that the original zero-mode flat band develops into valley-polarized current pathways.

Regarding the non-uniform magnetic field profiles, we have shown that they can be engineered to enhanced proximity induced effects in the band structure. It was also shown that the $k-$dependent coupling of TES can be explained in terms of the evolution of the magnetic confinement potential. For instance, we have seen that the non-uniform magnetic field can be defined to produced anti-chirality by allowing it to share symmetries with the magnetic field that would produce the anti-chiral modified Haldane model (AHM). This picture magnetic confining potentials can be exploited in future research to engineered one-dimensional edge states of arbitrary shapes and width by designing specific non-uniform magnetic field profiles.

## Acknowledgements

This work was supported by UNAM-PAPIIT IN116025. Ricardo Y. Diaz-Bonifaz thanks SECIHTI (formerly CONAHCYT) for the graduate scholarship granted. Computations were performed at Miztli under project LANCAD-UNAM-DGTIC-329.

## Appendix A. Peierls phase assignation for non-uniform magnetic fields

Magnetic fields can be introduced into tight-binding Hamiltonians by writing the hopping parameter between the $i-$th and $j-$th lattice site as $t_{i,j} = t_0 \exp(i\theta_{i,j})$. The terms $\theta_{i,j}$, known as Peierls phases (PPs), are commonly calculated by performing a line integral of the magnetic vector potential $\vec{A}$ between the $i-$th and $j-$th sites [45]. The gauge selection for $\vec{A}$ should allow the PPs to be identical for any unit cell of the ZGNR to enable the use of Bloch theorem, which is not straightforward for non-uniform magnetic fields. However, gauge freedom enables the use of a simple graphic algorithm to assign the phases conveniently using the magnetic flux per unit cell as the only constriction. This algorithm consists of assigning PPs sequentially for every bond in the lattice. The phases can be assigned at convenience, except when the bond forms a closed loop or cycle where all the phases have been previously assigned. In that case, the missing PP is calculated from the magnetic flux $\Phi_{cell}$ enclosed by the cycle and the previously assigned PPs $\theta_n$ along the closed path as

$$\theta_N = 2\pi \frac{\Phi_{cell}}{\Phi_0} - \sum_{n=1}^{N-1} \theta_n, \tag{8}$$

where $\theta_N$ is the only missing PP of a cycle formed by $N$ lattice sites and $\Phi_0 = h/e$. Further details and proof of this algorithm can be found in Ref. [39].

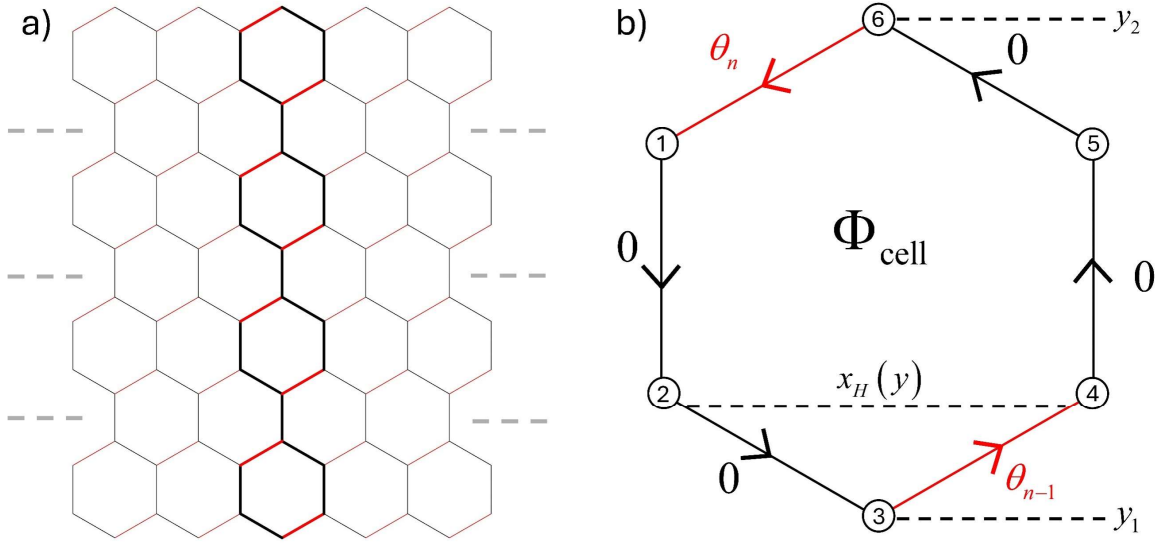

**Figure 8:** a) Representation of a portion of a ZGNR where a unit cell is highlighted. b) Schematic of a single hexagon within a ZGNR unit cell. The bonds for which the PP is non-zero are shown in red and $x_H(y)$ is the hexagon extension in the $x-$direction as a function of $y$.

As the magnetic fields hereby considered are uniform in the $x-$direction and vary exclusively along the $y-$direction, a ZGNR can be described using Bloch theorem as long as the PP assignation is identical for every unit cell of the ribbon. To this end, we propose a PP assignation as the one shown in Fig. 8, where the red lines indicate the bonds for which the PP is non-zero. Let us now consider a single hexagon from the ZGNR unit cell, as shown in Fig. 8b. If the phase $\theta_{n-1}$ has been previously assigned, the missing PP $\theta_n$ can be calculated from Eq. (8) as $\theta_n = 2\pi\Phi_{cell}/\Phi_0 - \theta_{n-1}$. As the magnetic field $B_z$ depends exclusively on $y$, the magnetic flux across the hexagon can be calculated as

$$\Phi_{cell} = \int_{y_1}^{y_2} B_z(y) x_H(y) dy, \tag{9}$$

where $x_H(y)$ is the hexagon width in the $x-$direction as a function of $y$. This procedure can be repeated to assign sequentially the PPs starting from the lower border of the ZGNR and going into the upper border, resulting in an identical phase set for every unit cell of the infinite nanoribbon.

**Appendix B. Equivalent magnetic field in modified Haldane model**

We now discuss how the Haldane model [28] and the anti-chiral modified Haldane model (AHM) [27] can be interpreted in terms of non-uniform magnetic fields. Both models can be written through a graphene TB Hamiltonian of the form

$$\hat{H} = t_0 \sum_{\langle n,m \rangle} \hat{c}_n^\dagger \hat{c}_m + t_1 \sum_{\langle\langle n,m \rangle\rangle} e^{-i\nu_{n,m}\Phi} \hat{c}_n^\dagger \hat{c}_m, \quad (10)$$

Where the first sum runs over nearest-neighbors (NN) and the second sum involves next-nearest-neighbors (NNN). Notice that the NNN sum contains complex terms $\exp(-i\nu_{n,m}\Phi)$, where $\Phi$ is a phase factor and $\nu_{n,m} = \pm 1$ is a sub-lattice specific factor. For the original Haldane model, $\nu_{n,m} = 1$ for sublattice A and $\nu_{n,m} = -1$ for sublattice B, which means that the phases acquired when drawing a clockwise loop of a single sublattice type (sublattice triangle) within a unit cell are all positive. In the AHM, $\nu_{n,m} = 1$ for both sublattices, which means that the phases acquired in a sublattice triangle are positive for one sublattice and negative for the other. A visual representation of phase acquisition can be found in Fig. 9. These complex phases break time reversal symmetry and are mathematically indistinguishable from Peierls phases introduced by an external magnetic field. In the following we address the properties and symmetries that a magnetic field must possess in order to obtain the Hamiltonian (10).

Let us start by mentioning that the net magnetic flux through any unit cell must be zero, as the sum of phases over a hexagon border following nearest-neighbor bonds is zero. However, notice that if we draw a triangle within a unit cell whose vertices are from the same sublattice, the sum of phases is $3\nu_{n,m}\Phi$, which means that there is a non-zero magnetic flux enclosed by the triangle. In the Haldane model, the magnetic flux enclosed by both sublattices is identical but in the modified Haldane model the magnetic flux changes sign within sublattices. This loss of sublattice symmetry is used by Colomés and Franz to offset the Dirac points of a ZGNR nanoribbon and produce anti-chirality [27]. Nonetheless, it is worth noting that for the effect to be perceived, the borders might be sublattice polarized to a degree. For instance, in a ZGNR both borders are formed from different sublattices, which enhances the effects that rely on the absence of this symmetry. Conversely, an armchair graphene nanoribbon (AGNR) is not sublattice polarized and the effect is lost. A proof of this relies in the fact that in an AGNR there is a single Dirac point [36], so no offset between sublattices can produce a TES. This means that the AHM depends on the border type to produce anti-chirality and the edge states that appear are not robust to changes in the border orientation.

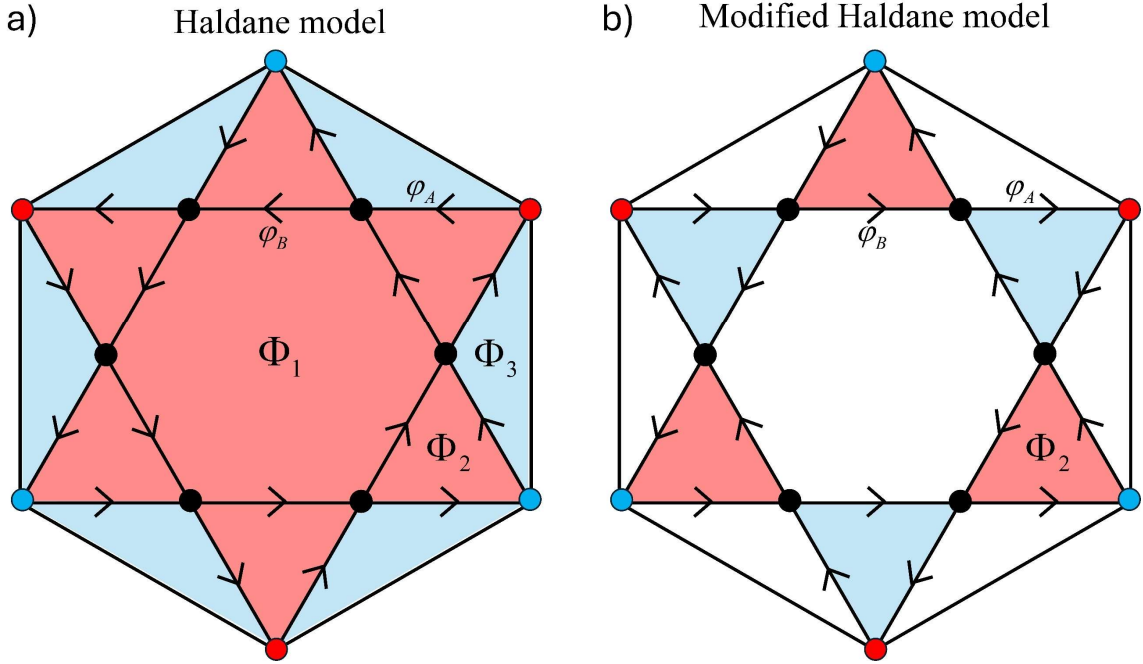

**Figure 9:** Diagram of the unit cells in the Haldane (a) and modified Haldane (b) models. In both cases, the arrows indicate the direction in which the acquired phase is positive. The red and blue dots denote sites from sublattices A and B, while the black dots are not lattice sites but intersections between NNN bonds. The color of each area denotes if the local magnetic flux is positive (red), negative (blue) or zero (white).

A magnetic field that reproduces the AHM is not unique. To better understand the magnetic flux distribution over the unit cell we can think of the different areas delimited by the bonds. This is shown in Fig. 9, where the red and blue dots are sites from sublattice A and B respectively. The black dots are intersections between NNN bonds which are not lattice sites but are useful to define sections of the unit cell. One way of defining the magnetic field is by assuming that an electron acquires the phase uniformly as it travels from one site to their NNN. By assuming this, the total phase of a NNN hopping parameter can be written as $\Phi = 2\varphi_A + \varphi_B$, where $\varphi_A$ is the phase acquired when traveling from a lattice site to a black dot and $\varphi_B$ is the phase acquired when traveling between black dots. Within this description we can see that a field that produces the Haldane model has the symmetries of the dihedral group symmetries $D_6$. However, the magnetic fluxes in the AHM constrain the symmetry transformations to the dihedral group $D_3$. In particular, a ZGNR described by the AHM is not invariant under reflections over the axis in which it is periodic, making the upper and lower edges inequivalent as each one can be associated with opposite local magnetic fields.

**References**

[1] F. Giustino et al., The 2021 quantum materials roadmap, J. Phys. Mater. **3**, 042006 (2021).

[2] M. El-Batanouny, *Advanced Quantum Condensed Matter Physics: One-Body, Many-Body, and Topological Perspectives* (Cambridge University Press, Cambridge, 2020).

[3] S. Yin, L. Ye, H. He, X. Huang, M. Ke, W. Deng, J. Lu, and Z. Liu, Valley edge states as bound states in the continuum, Science Bulletin **69**, 1660 (2024).

[4] L. Ju et al., Topological valley transport at bilayer graphene domain walls, Nature **520**, 650 (2015).

[5] T. Hou, Y. Ren, Y. Quan, J. Jung, W. Ren, and Z. Qiao, Metallic network of topological domain walls, Phys. Rev. B **101**, 201403 (2020).

[6] S. G. y García, Y. Betancur-Ocampo, F. Sánchez-Ochoa, and T. Stegmann, Atomically Thin Current Pathways in Graphene through Kekulé-O Engineering, Nano Lett. **24**, 2322 (2024).

[7] G. W. Semenoff, V. Semenoff, and F. Zhou, Domain Walls in Gapped Graphene, Phys. Rev. Lett. **101**, 087204 (2008).

[8] K. V. Klitzing, G. Dorda, and M. Pepper, New Method for High-Accuracy Determination of the Fine-Structure Constant Based on Quantized Hall Resistance, Phys. Rev. Lett. **45**, 494 (1980).

[9] D. J. Thouless, M. Kohmoto, M. P. Nightingale, and M. den Nijs, Quantized Hall Conductance in a Two-Dimensional Periodic Potential, Phys. Rev. Lett. **49**, 405 (1982).

[10] M. Kohmoto, Topological invariant and the quantization of the Hall conductance, Annals of Physics **160**, 343 (1985).

[11] P. Carmier, C. Lewenkopf, and D. Ullmo, Graphene $n\text{-}p$ junction in a strong magnetic field: A semiclassical study, Phys. Rev. B **81**, 241406 (2010).

[12] J. R. Williams, L. DiCarlo, and C. M. Marcus, Quantum hall effect in a gate-controlled p-n junction of graphene, Science **317**, 638 (2007).

[13] P. Carmier, C. Lewenkopf, and D. Ullmo, Semiclassical magnetotransport in graphene $n$-$p$ junctions, Phys. Rev. B **84**, 195428 (2011).

[14] D. A. Abanin and L. S. Levitov, Quantized Transport in Graphene p-n Junctions in a Magnetic Field, Science **317**, 641 (2007).

[15] C.-Z. Chang, C.-X. Liu, and A. H. MacDonald, *Colloquium*: Quantum anomalous Hall effect, Rev. Mod. Phys. **95**, 011002 (2023).

[16] Y.-F. Zhao, R. Zhang, J. Cai, D. Zhuo, L.-J. Zhou, Z.-J. Yan, M. H. W. Chan, X. Xu, and C.-Z. Chang, Creation of chiral interface channels for quantized transport in magnetic topological insulator multilayer heterostructures, Nat Commun **14**, 770 (2023).

[17] D. Ovchinnikov et al., Topological current divider in a Chern insulator junction, Nat Commun **13**, 5967 (2022).

[18] L.-J. Zhou et al., Confinement-Induced Chiral Edge Channel Interaction in Quantum Anomalous Hall Insulators, Phys. Rev. Lett. **130**, 086201 (2023).

[19] K. Yasuda, M. Mogi, R. Yoshimi, A. Tsukazaki, K. S. Takahashi, M. Kawasaki, F. Kagawa, and Y. Tokura, Quantized chiral edge conduction on domain walls of a magnetic topological insulator, Science **358**, 1311 (2017).

[20] Y. Han, S. Pan, and Z. Qiao, Topological junctions in high-Chern-number quantum anomalous Hall systems, Phys. Rev. B **108**, 115302 (2023).

[21] K. Novoselov, Mind the gap, Nature Mater **6**, 720 (2007).

[22] Y. Cao, V. Fatemi, S. Fang, K. Watanabe, T. Taniguchi, E. Kaxiras, and P. Jarillo-Herrero, Unconventional superconductivity in magic-angle graphene superlattices, Nature **556**, 43 (2018).

[23] J. M. Park, Y. Cao, K. Watanabe, T. Taniguchi, and P. Jarillo-Herrero, Tunable strongly coupled superconductivity in magic-angle twisted trilayer graphene, Nature **590**, 249 (2021).

[24] E. Andrade, F. López-Urías, and G. G. Naumis, Flat bands without twists through the use of a one harmonic Moiré systems: topological nature of modes and electron–electron pairing in periodic uniaxial strained or crenellated graphene nanoribbons, 2D Mater. **12**, 015016 (2024).

[25] P. Roman-Taboada and G. G. Naumis, Topological flat bands in time-periodically driven uniaxial strained graphene nanoribbons, Phys. Rev. B **95**, 115440 (2017).

[26] J. Jiang, Q. Gao, Z. Zhou, C. Shen, M. Di Luca, E. Hajigeorgiou, K. Watanabe, T. Taniguchi, and M. Banerjee, Direct probing of energy gaps and bandwidth in gate-tunable flat band graphene systems, Nat Commun **16**, 1308 (2025).

[27] E. Colomés and M. Franz, Antichiral Edge States in a Modified Haldane Nanoribbon, Phys. Rev. Lett. **120**, 086603 (2018).

[28] F. D. M. Haldane, Model for a Quantum Hall Effect without Landau Levels: Condensed-Matter Realization of the "Parity Anomaly," Phys. Rev. Lett. **61**, 2015 (1988).

[29] M. Yan, X. Huang, J. Wu, W. Deng, J. Lu, and Z. Liu, Antichirality Emergent in Type-II Weyl Phononic Crystals, Phys. Rev. Lett. **130**, 266304 (2023).

[30] Y. Yang, D. Zhu, Z. Hang, and Y. Chong, Observation of antichiral edge states in a circuit lattice, Sci. China Phys. Mech. Astron. **64**, 257011 (2021).

[31] C. L. Kane and E. J. Mele, Quantum Spin Hall Effect in Graphene, Phys. Rev. Lett. **95**, 226801 (2005).

[32] M. König, S. Wiedmann, C. Brüne, A. Roth, H. Buhmann, L. W. Molenkamp, X.-L. Qi, and S.-C. Zhang, Quantum Spin Hall Insulator State in HgTe Quantum Wells, Science **318**, 766 (2007).

[33] D. R. Hofstadter, Energy levels and wave functions of Bloch electrons in rational and irrational magnetic fields, Phys. Rev. B **14**, 2239 (1976).

[34] R. Rammal, Landau level spectrum of Bloch electrons in a honeycomb lattice, J. Phys. France **46**, 1345 (1985).

[35] J. E. Avron, D. Osadchy, and R. Seiler, A Topological Look at the Quantum Hall Effect, Physics Today **56**, 38 (2003).

[36] A. H. Castro Neto, F. Guinea, N. M. R. Peres, K. S. Novoselov, and A. K. Geim, The electronic properties of graphene, Rev. Mod. Phys. **81**, 109 (2009).

[37] L. E. F. Foa Torres, S. Roche, and J.-C. Charlier, *Introduction to Graphene-Based Nanomaterials: From Electronic Structure to Quantum Transport* (Cambridge University Press, Cambridge, 2014).

[38] Y. Aharonov and D. Bohm, Significance of Electromagnetic Potentials in the Quantum Theory, Phys. Rev. **115**, 485 (1959).

[39] R. Y. Díaz-Bonifaz and C. Ramírez, Dots and boxes algorithm for Peierls substitution: application to multidomain topological insulators, J. Phys.: Condens. Matter **37**, 105301 (2025).

[40] B. Kramer and A. MacKinnon, Localization: theory and experiment, Rep. Prog. Phys. **56**, 1469 (1993).

[41] C. Ramírez, F. H. González, and C. G. Galván, Solving Schrödinger Equation with Scattering Matrices. Bound States of Lennard-Jones Potential, J. Phys. Soc. Jpn. **88**, 094002 (2019).

[42] L. D. Landau and L. M. Lifshitz, *Quantum Mechanics: Non-Relativistic Theory*, 3rd ed. (Butterworth-Heinemann, 1981).

[43] T. B. Boykin, M. Luisier, and G. Klimeck, Current density and continuity in discretized models, Eur. J. Phys. **31**, 1077 (2010).

[44] X.-L. Lü, J.-E. Yang, and H. Chen, Manipulation of antichiral edge state based on modified Haldane model, New J. Phys. **24**, 103021 (2022).

[45] A. Cresti, Convenient Peierls phase choice for periodic atomistic systems under magnetic field, Phys. Rev. B **103**, 045402 (2021).